\documentclass[default]{svmult}

\usepackage{type1cm}        % activate if the above 3 fonts are
\usepackage{makeidx}         % allows index generation
\usepackage{graphicx}        % standard LaTeX graphics tool
\usepackage{multicol}        % used for the two-column index
\usepackage[bottom]{footmisc}% places footnotes at page bottom

\usepackage{newtxtext}       % 
\usepackage[varvw]{newtxmath}       % selects Times Roman as basic font

\usepackage{comment}
\usepackage{aas_macros}     % AAS shortcuts
\usepackage{wrapfig}

\usepackage[breaklinks,colorlinks,urlcolor=blue,citecolor=blue,linkcolor=magenta]{hyperref}

\newcommand{\be}{\begin{equation}} 
\newcommand{\ee}{\end{equation}}
\newcommand{\bea}{\begin{equation}\begin{aligned}} 
\newcommand{\eea}{\end{aligned}\end{equation}}
\usepackage[version=4]{mhchem}
\usepackage{siunitx}
\DeclareSIUnit\year{yr}

\newcommand\mpl{
    M_{\mathrm{Pl}}
}

\newcommand\du{
    \mathrm{d}
}
\newcommand\dd{
    \,\du
}

\def\lsim{\mathrel{\raise.3ex\hbox{$<$\kern-.75em\lower1ex\hbox{$\sim$}}}}
\def\gsim{\mathrel{\raise.3ex\hbox{$>$\kern-.75em\lower1ex\hbox{$\sim$}}}}

\makeindex             % used for the subject index
\begin{document}

% -----------------------------------------------------------------------------

\graphicspath{{Figures/}}

\title*{Ultralight Primordial Black Holes}
\author{S. Profumo}%\orcidID{0000-0002-9159-7556}}
\institute{
\at University of California, Santa Cruz, and Santa Cruz Institute for Particle Physics\\ 1156 High St, Santa Cruz, CA 95060 U.S.A.
 \email{profumo@ucsc.edu}}

\maketitle%\label{chapter:GWmergers}

% -----------------------------------------------------------------------------

\abstract{The fate of ultralight black holes depends on whether or not evaporation stops at or around the Planck scale. If evaporation stops, the general expectation is that a population of Planck-scale will be left over, possibly including a significant fraction of electrically charged relics. If evaporation does not stop, a runaway ``explosion'' would occur, with significant and potentially detectable high-energy emission. Here, I review both possibilities, with an emphasis on current status and future detection prospects.}

%{\bf Structure of the chapter:}
%\begin{itemize}
%    \item 
%\end{itemize}

\section{Introduction}
Primordial black holes (PBHs) produced in the very early universe with mass below, approximately, $M_U\simeq 5.1\times 10^{14}$ g (absent dark degrees of freedom, that could significantly shorten the timescale for evaporation \cite{Baker:2022rkn}) have completed evaporation. The end products of evaporation are highly uncertain, as when $M_{\rm PBH}\to M_{\rm Pl}$ the BH temperature $T_{\rm BH}\to M_{\rm Pl}$ and, thus, quantum gravity effects are expected to play an important role \cite{Carr:2020gox}. Broadly, evaporation could in principle continue at sub-Planckian scales (corresponding to trans-Planckian temperatures), eventually consuming the BH, and $M_{\rm BH}\to 0$, or stop around, or below the Planck scale, leaving a stable relic. In addition, BHs can approach extremality as $M_{\rm PBH}\to M_{\rm Pl}$, which could also result in the end of the evaporation process \cite{Lehmann_2019}.

In this chapter I discuss the end-point of PBH evaporation: In the following sec.~\ref{sec:Plrelics} I review the possibility of stable Planck-scale relics, possibly, but not necessarily constituting a fraction, or all, of the cosmological dark matter. The following sec.~\ref{sec:Pldetection} then details how such Planck-scale relics can be directly detected with an array of experimental facilities. I then turn, in sec.~\ref{sec:explosions} to the opposite possibility of complete evaporation and of the direct detection of the runaway high-energy emission from PBHs exploding today. The final sec.~\ref{sec:conclusions} summarizes and concludes.

\section{Planck-scale relics}\label{sec:Plrelics}  

%how they originate
The notion that Planck-scale relics of PBH evaporation could be some, or all, the cosmological dark matter (DM) has been contemplated for a long time (see e.g. \cite{MacGibbon:1987my}). As a BH mass approaches, its mass depleted by increasingly fast Hawking evaporation, the Planck scale, one out of three scenarios is conjectured to occur:\\ 

(i) a stable, Planck-scale relic is left over \cite{MacGibbon:1987my};\\ 

(ii) a naked spacetime singularity if left over \cite{DeWitt:1975ys};\\ 

(iii) evaporation continues at trans-Planckian energies and temperatures, and the singularity vanishes \cite{Hawking:1976ja}.\\ 

Any PBH formed in the early universe with a lifetime shorter than the age of the universe (as mentioned above corresponding to a mass at production of $M_U$) is slated to fall into one of these three categories.

%whether or not they are charged at late times
In this section, I will assume that (i) or (ii) is the case. General arguments then indicate that Planck-scale relics can be either electromagnetically neutral, or carry a residual electric or magnetic charge\footnote{In principle dark sector charges are also possible.}. The experimental absence of magnetic monopoles does not necessarily exclude the possibility of magnetically charged Planck relics, although, admittedly, this possibility is rather exotic \cite{Diamond:2021scl}. I will thus entertain the possibility that a fraction $f_{\rm CPR}$ of relics be electrically charged. Absent a charge, Planck scale relic are virtually entirely undetectable. Also note that there are general arguments \cite{Page:1976ki} for why even if PBHs were produced with significant initial spin parameter $a=J/M$, with $J$ the PBH's angular momentum and $M$ its mass, evaporation rapidly and efficiently sheds angular momentum, leaving $a\to0$ at the end of evaporation for $M<M_U$.

%they are not too fast, unless...
Note that Planck relics may have a significant initial velocity, as their momentum results from the ejection of highly energetic quanta which, around the Planck scale, have energies on the order of the Planck scale itself. The resulting relic's momentum is effectively a random walk in momentum space, dominated by the emission at the end of evaporation, when the Hawking temperature is highest. For $N_q$ terminal emitted quanta, the momentum of the BH goes as $1/\sqrt{N_q}$. Thus, at most, calling $M_R$ the relic mass, the relic's momentum $p\simeq M_R$ (corresponding to $N_q=1$) and the energy $E=\sqrt{2}M_R$, thus the Lorentz factor of the relic at ``production'' is, at most, $\gamma\simeq \sqrt{2}$ and the average initial velocity $\bar v_i\simeq 1/\sqrt{2}$. This corresponds to a momentum, today, $p_0=a_ip_i$, where $a_i$ is the scale factor at production. Structure formation limits the cosmological dark matter velocity today to be $\lesssim 5\times 10^{-7}$ \cite{Fujita:2014hha,Morrison:2018xla,Viel:2005qj} (see also \cite{Lehmann:2021ijf}), which, in turn, means that any relic produced before BBN (when the scale factor $a_{\rm BBN}\simeq 2.5\times 10^{-10}$), as it should given the strong constraints on evaporation after BBN, is highly non-relativisitc \cite{Lehmann:2021ijf}.

Whether or not Planck relics are stable against discharge is a matter of some controversy. There are several reasons why a first-principles computation of the relic charge on Planck relics is problematic. First, it is unclear whether one can neglect the back-reaction of the evaporation products on the BH, and, as a consequence, whether or not the particle emission can be considered as an ``individual'', isolated event \cite{Page:1976ki}; Second, the charge-to-mass ratio may get large enough that it could significantly affect, or even stop, the evaporation rate: recall that a Planck-mass BH reaches extremality for $Q\simeq 12\ e$. Third, it is unclear how the running of the electromagnetic coupling $\alpha_{\rm EM}$ behaves at or near the Planck scale, where several layers of new physics possibly come into play through radiative effects.

These conceptual difficulties non-withstanding, one can proceed to an analytical estimate of the relic charge on a Planck relic whose evaporation stops at some mass scale $M_R\simeq M_{\rm Pl}$ (in what follows I always assume for definiteness $M_R= M_{\rm Pl}$). Page \cite{Page:1977um} estimated that the probability distribution for BH charges $Q$ is approximately Gaussian,
\begin{equation}
    P(Q)\sim\exp\left(-4\pi\alpha(Q/e)^2\right),
\end{equation}
with an rms value $Q/e=1/\sqrt{8\pi\alpha}\simeq 2.34$. However, if the product of the BH mass and the emitted charged particle mass is small enough in Planck units, then the rms value $Q/e$ increases up to approximately 6.

The more modern approach of Ref.~\cite{Lehmann_2019} relaxes some assumption of the original Page calculation, such as taking into account the finite value of the BH mass times the light lepton mass, and without neglecting the BH charge in the computation of the grey body factors. The results are qualitatively similar to the original estimate, and the fraction of charged to neutral relics was found to be of order 1.

The fate of the charged relics depends on their cosmic history. It is easy to realize that direct accretion of opposite-sign particles would not discharge the BH, since the accretion cross section is approximately geometric, thus $\sigma\sim M_R^2/M_{\rm Pl}^4\sim 1/M_{\rm Pl}^2$ and the accretion rate $\Gamma\lesssim n_e/ M_{\rm Pl}^2$ is always much smaller than the Hubble rate $H\sim T^2/M_{\rm Pl}$, with $T$ the universe's temperature, since 
\begin{equation}
\frac{\Gamma(T)}{H(T)}\sim \frac{T^3}{M_{\rm Pl}^2}\frac{M_{\rm Pl}}{T^2}\sim \frac{T}{M_{\rm Pl}}\ll1\ {\rm for}\  T\ll M_{\rm Pl}.
\end{equation}
Much more likely is the possibility that charged, Planck-scale relics (CPR) form bound states with opposite-sign cosmic rays. In this case, however, reionization proceeds significantly quicker \cite{1964ApJS....9..185B} than for visible-sector particles \cite{Lehmann_2019}, and thus, today, CPRs are not expected to exist in neutral bound states.

\section{Direct detection of Charged Planck-scale relics}\label{sec:Pldetection}

%overview
CPRs are extremely heavy, resulting in a small direct detection event rate $N$, which for nominal values of the local DM density and velocity, and for a detector with  effective area $A_{\rm eff}$ and efficiency $\varepsilon$ reads
\begin{equation}
    N\simeq 0.23\ {\rm yr}^{-1} f_{\rm CPR}\left(\frac{M_R}{M_{\rm Pl}}\right)^{-1}\left(\frac{\rho_{\rm DM}}{0.3\ {\rm GeV/cm}^3}\right)\left(\frac{v_{\rm DM}}{300\ {\rm km/s}}\right)\left(\frac{A_{\rm eff}}{1\ m^2}\right)\varepsilon.
\end{equation}
CPRs would behave in particle detectors similarly to very heavy ions, with kinetic energies on the order of 
\begin{equation}
    K\sim\frac{1}{2}M_R v_{\rm DM}^2\simeq 6\times 10^{16}\ {\rm eV}\ \ (M_R\simeq M_{\rm Pl})
\end{equation}
This should be compared with typical atomic energy scales in the 1-$10^3$ eV: CPRs  travel through matter without significantly decreasing their kinetic energy, and essentially with no deflection, as the relative momentum transfer is also negligible.

Most of the energy deposited by the passage of a CPR through matter is delivered to nuclei and not to electrons, since $m_e\ll m_N$ and $\Delta E\sim m v_{\rm DM}^2$. Nuclear scattering of course will also produce ionization and scintillation, albeit with target-dependent quenching factors \cite{Tretyak:2009sr,Mu:2013pja}. Interestingly, Ref.~\cite{Lazanu:2020qod} finds that the signals expected for instance in liguid Argon, ionization and scintillation, by CPRs would enable the discrimination between those tracks and those expected from other dark matter candidates. In particular, the CPRs
 trajectories are uniquely slated to appear as crossing the whole active medium, in every direction,
producing uniform ionization and scintillation on the whole path \cite{Lazanu:2020qod}.

\begin{figure}
    \centering
    \includegraphics[width=0.95\columnwidth]{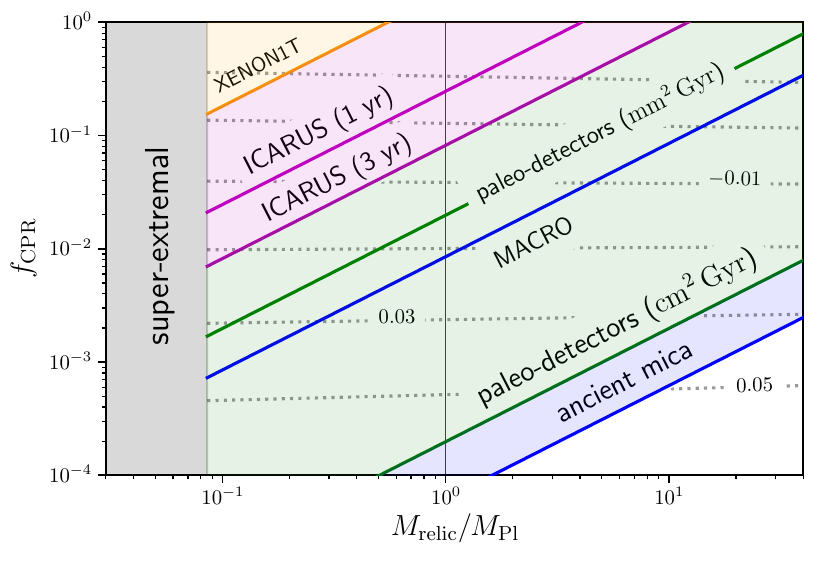}
    \caption{
        Projected 99\% CL upper limit on the mass and density of CPRs with experiments of several classes. See text for details. \textbf{Orange:} a 10 yr exposure of XENON1T \cite{Aprile:2018dbl}. \textbf{Magenta:} solid: a 3 yr  exposure of ICARUS. The dashed line shows a 1 yr  exposure. \textbf{Green:} solid: estimated limits from a paleo-detector with $\mathcal E=1$ and a 1 cm$^2$ Gyr exposure. The dashed line shows a 1 mm$^2$ Gyr exposure. \textbf{Blue:} strongest possible limits from monopole searches, including a direct search by MACRO and a search for tracks in ancient mica \cite{Ghosh_1990}. \textbf{Dotted gray:} relic fractions produced assuming an initial a power-law mass function with index $\gamma$. Contours step from $\gamma=-0.05$ to $\gamma=0.05$ from top to bottom in increments of $0.02$. \textbf{Shaded gray:} region prohibited by super-extremality for a charge of $1e$.  Figure reproduced, with permission, from Ref.~\cite{Lehmann_2019}.
    }
    \label{fig:constraints}
\end{figure}

Several classes of detectors are suitable to search for CPRs. Bubble chambers of superheated fluids such as PICO \cite{PICO:2016kso}  are sensitive to the passage of a highly ionizing massive CPR, according to the simulations in \cite{Lehmann_2019}, leaving a signal significantly different from that of both WIMPs and neutron bakcgrounds. Unforunately, the event rate for the proposed 500 L version of PICO is too small to place significant constraints on CPRs (see fig.~\ref{fig:constraints}.

Ref.~\cite{Lehmann_2019} estimates that the photon yield at atmsopheric fluorescence detectors such as HiRes \cite{HiRes:2004hvo} would produce a signal to noise too small to be detectable, even with very favorable assumptions on the detector's fluorescence efficiency. Similar issues apply to circumventing trigger issues in Cherenkov detectors, given that the CPRs are relatively slow.

Direct dark matter detectors are, on the other hand, ideally placed to detect CPRs: the overburdain is completely irrelevant in stopping the CPRs, and some of the largest noble gas detectors have a significant effective area and very high efficiency. Finally, so-called paleo-detectors have such long effective exposure times that they provide the best opportunity for the direct detection of CPRs.

%constraints
Fig.~\ref{fig:constraints} (reproduced, with permission, from Ref.~\cite{Lehmann_2019}) shows  projected, future constraints from a variety of experimental facilities. In particular, the left, grey region corresponds to masses for which the PBHs would be super-extremal for a charge $Q=1\ e$, and the vertical line corresponds to $M_R=M_{\rm Pl}$. Shaded regions are testable, or ruled out, by direct dark matter search experiments such as XENON1T (orange), and by neutrino experiments such as ICARUS (magenta). The green region corresponds to constraints from paleodetectors with different effective exposures. Finally, the blue region is constrained by monopole searches with MACRO and for tracks in ancient mica \cite{Ghosh_1990}. 

The dashed lines reflect the possibility that the PBHs be initially distributed according to a power-law initial mass function $M \dd N/\du M\propto M^{\gamma-1}$. Assuming that the entirety of the cosmological dark matter is in the form of PBHs, and that black holes with mass $M>M_U$ do not lose any significant amount of mass, 
the mass fraction in CPRs is for $\gamma<0$, 
\begin{equation}
    \frac{\Omega_{\mathrm{CPR}}}{\Omega_{\mathrm{DM}}} \approx \frac{
        \mpl M_{U}^{\gamma-1} - \mpl^{\gamma}
    }{
        M_{U}^{\gamma-1}\left[
            \mpl - (1-1/\gamma)M_{U}
        \right] - \mpl^\gamma
    }.
\end{equation}
A CPR fraction $f\sim1$ is produced when $\gamma\lesssim-0.1$, whereas e.g. $f\sim10^{-2}$ for $\gamma\sim-10^{-2}$. Fig.~\ref{fig:constraints} shows contours corresponding to $\gamma=-0.05$ up to 0.05 in increments of 0.02.

A final, interesting, labeit rather speculative, possibility is that Planck-scale relics, whether charged or not, could merge and decay, emitting Planck-scale cosmic and gamma rays that could be detectable, in principle, with very high-energy cosmic ray telescopes \cite{Barrau:2019cuo}

\section{Exploding Black Holes}\label{sec:explosions}
%In the early universe: gravity waves, baryogenesis
The evaporation of light PBH in the early universe is tightly constrained if happening after Big Bang Nucleosynthesis (BBN) \cite{Carr:2020gox}. The evaporation products of PBHs evaporating {\em prior} to BBN, on the other hand, are expected to quickly thermalize, leaving no measurable imprint, with at least three important exceptions:
\begin{enumerate}
    \item If the PBH energy density is comparable, or larger than the energy density of the rest of the universe, then the expansion rate is affected by evaporation, that acts as an episode of ``reheating''; constraints can thus arise if such process is close enough to BBN;\\
    
    \item PBH evaporation prior to BBN can, in principle, be responsible for the production of dark sector species \cite{Morrison:2018xla}, including the cosmological dark matter, and even of the baryon asymmetry (for instance via the production of right-handed heavy neutrinos whose decay is CP violating, as in the standard leptogenesis scenario \cite{Pilaftsis:2009pk}, or via effective operators that are CP violating and couple to the $B-L$ current, as envisioned in \cite{Smyth:2021lkn};\\
    
    \item An evaporation product that is slated not to thermalize is gravitons: graviton production would manifest itself as a stochastic background of very high-frequency gravitational waves \cite{Dolgov:2011cq}. The frequency of the signal could be lowered in principle by either a non-standard cosmological history in the early universe \cite{Ireland:2023avg}, or by the existence of extra-dimensions, with an associated suppressed Planck scale \cite{Ireland:2023zrd}.
\end{enumerate}

%Now, produced in the early universe
PBHs produced in the early universe may, in principle, be at the end of their evaporation process today. Ref.~\cite{Boluna:2023jlo} calculates that for an initial mass function $\psi(M)\equiv MdN/dM$, the rate of PBH evaporating today is
\begin{equation}
    \dot n_{\rm PBH}\simeq\rho_{\rm DM}\frac{\psi_i\left(M_U \right)}{3t_U},
\end{equation}
with $t_U$ the age of the universe. Depending on the mass function, constraints on the PBH abundance bound this rate to be generically rather low, at most around one event per cubic parsec per year, in the most favorable possible case of a very narrow lognormal mass function. This should be compared with the HAWC limits on the explosion rate, which at present is $\dot n_{\rm PBH}<3,400\ {\rm pc}^{-3}{\rm yr}^{-1}$.

%produced now
On the other hand, PBHs could arise in the {\em late} universe (see the discussion in Ch.~10 of this book). If that is the case, the rate of explosions is virtually unconstrained, and could be very large. As a result, PBHs could be associated with astroparticle anomalies such as the Galactic center gamma-ray excess \cite{Fermi-LAT:2017opo}, or that, as well as the AMS tentative $^3$He events \cite{Carlson:2014ssa,Coogan:2017pwt}, and the antiproton excess \cite{Cui:2016ppb}.

%how to detect them
PBH ``explosions'' (as Hawking referred to them \cite{Hawking:1974rv}) look like ``backwards'' gamma-ray bursts: the luminosity and photon energy grows in a runaway process, which suddenly stops as evaporation completes. Ref.~\cite{Boluna:2023jlo} searched for such events both in the Fermi GRB catalogue \cite{vonKienlin:2020xvz} and in the Fermi Large Area Telescope transient sources \cite{Fermi-LAT:2021ykq}, in the case where evaporation may not have completed yet. Critically, any events potentially associated with exploding PBHs would most likely have a non-trivial proper motion in the sky, which was not observed for any of the Fermi candidate events. As a result, there is no current evidence for any PBH explosion, consistent with theoretical expectations. 

\begin{figure}[!t]
    %\centering
    \includegraphics[width=\linewidth]{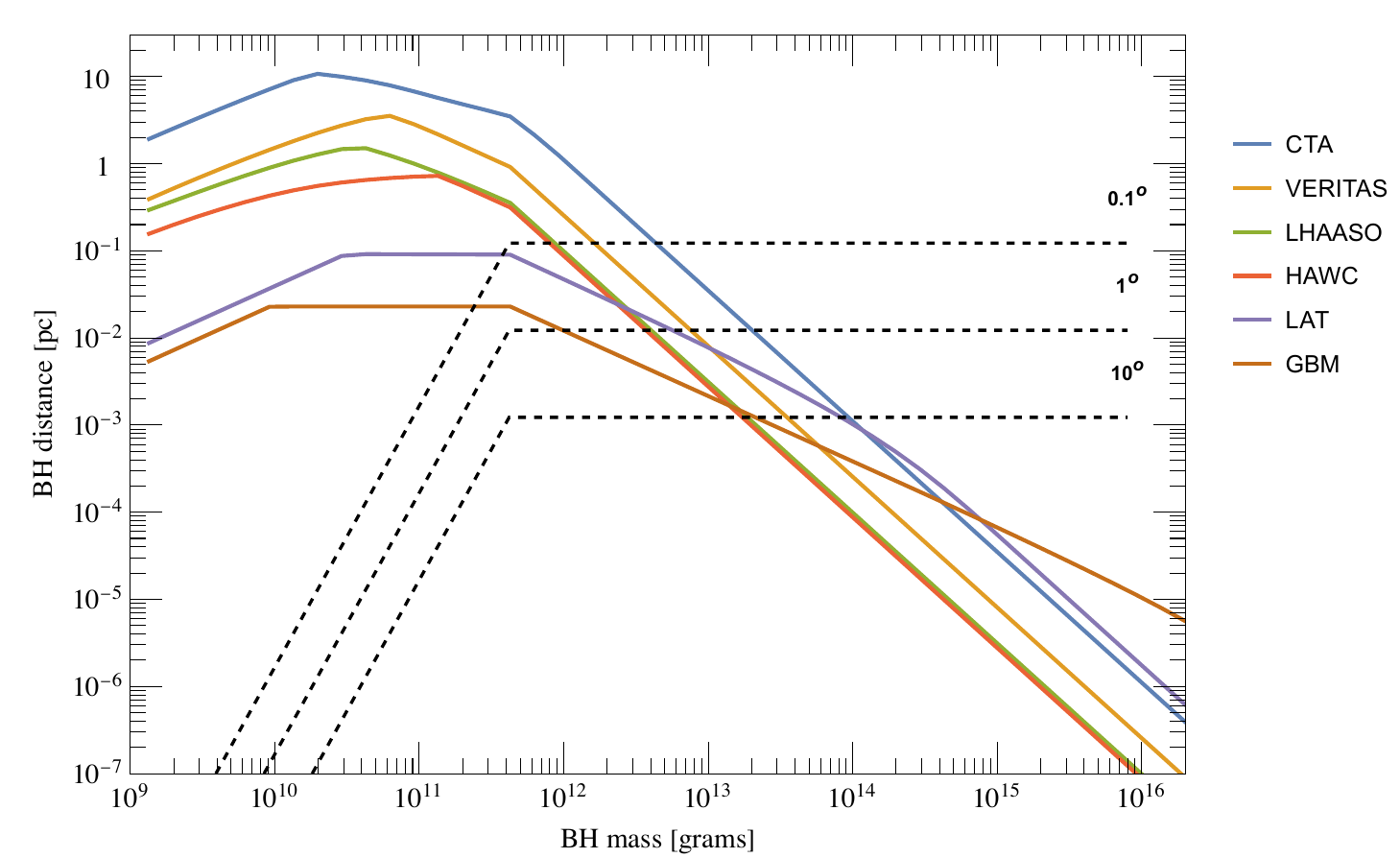}
    \caption{Maximum visible distance curves with respect to black hole mass for a number of modern gamma-ray detectors. The black dashed lines indicate the distance, at a given mass, at which the  proper motion during an observation up to one year long may exceed 0.1, 1 and 10 degrees. Figure reproduced, with permission, from Ref.~\cite{Boluna:2023jlo}.}
    \label{fig:detsens}
\end{figure}

Ref.~\cite{Boluna:2023jlo}'s results, summarized in fig.~\ref{fig:detsens}, reproduced with permission from \cite{Boluna:2023jlo}, indicate that even the most competitive future telescope, the Cherenkov Telescope Array (CTA) \cite{CTAConsortium:2013ofs} would only be able to observe an exploding PBH if it were closer than, approximately, 1 pc, under the best possible circumstances. As such, it may be possible to actually measure time of arrival differences with satellites positioned at macroscopic distances from one another: 
Observations with the Interplanetary Gamma-Ray Burst Timing Network\footnote{\tt https://heasarc.gsfc.nasa.gov/docs/heasarc/missions/ipn.html} (IPN) of GRB detectors could likely be the most promising search technique to identify a relatively nearby GRB-like event from a genuine, extra-galactic GRB \cite{Ukwatta:2015mfb}. Searches for local events are currently under way.

Figure \ref{fig:detsens} also shows that a large proper motion (shown with black dashed lines) is expected to occur if the evaporation is not yet in its runaway, final phase. 
%
%IPN

%dark degrees of freedom
The observation of PBH explosions could very strongly constrain the existence of otherwise ``dark'' (meaning not sharing any interaction with visible sector particles) degrees of freedom. BH evaporation is independent of particle interactions, and as such, as soon as a particle is kinematically accessible to BH evaporation, BHs need evaporate into it, shortening the runaway evaporation/explosion process. High-energy gamma-ray telescopes such as HAWC \cite{HAWC:2011gts} and CTA \cite{CTAConsortium:2013ofs} are ideally suited to search for such dark degrees of freedom in the putative light curves of evaporating PBHs \cite{Baker:2022rkn}.

A staggering possible fate of light PBH evaporation is to form ``hot spots'' in the early universe that would lead to a certain, finite decay rate of metastable vacua, including the
electroweak vacuum; this could lead, in turn, and assuming the validity of the posited temperature profile of such hot spots, to significant constraints on the initial abundance of light PBH \cite{Hamaide:2023ayu}.

An additional interesting possibility is that PBH be responsible for reheating at the end of inflation, as discussed e.g. in Ref.~\cite{Hidalgo:2011fj}. In this scheme, ultralight PBHs are formed (e.g., from the inflaton field) after inflation, go on to dominate the energy density of the Universe, and evaporate before Big Bang Nuncleosynthesis begins. In this scenario non-gravitational couplings between the inflaton and the SM  are not required, making PBH reheating and the 
corresponding inflation scenarios rather economic from a model building perspective.

%\section{Detection of Black Hole Explosions}\label{sec:explosiondetection}

% ALSO MENTION scattering of Planck relics as a formation mechanism for PBH! https://journals.aps.org/prd/pdf/10.1103/PhysRevD.100.123505

\section{Conclusion and Outlook}\label{sec:conclusions}
The most likely fate of light PBHs is either complete evaporation in a runaway process, or the formation of Planck-scale relics. There exist several arguments for the latter possibility, but the first is not excluded, as physics at the Planck scale is largely unknown. Here, I reviewed how the existence of such light relics, or exploding PBHs, can be tested with experiments and observations.

Unless Planck-scale relics are electrically or magnetically charged, or unless they merge and evaporate in the late universe at appreciable rates, such particles are virtually undetectable. The most competitive detectors, for relic charged Planck-scale BHs, are paleo-detectors - ancient crystals whose effective exposure time is on the order of the age of the Earth, and that can be searched for for defects created by the passage of a highly-ionizing, massive particle. Neutrino experiments and large, noble-gas direct dark matter detectors are also rather competitive.

Exploding PBHs - either coming from an initial mass function serendipitously peaked at the mass corresponding to a BH lifetime approximately equal to the age of the universe today, or formed by some non-primordial mechanism at late times - can be detected searching for what is, effectively, a ``reverse gamma-ray burst'': different telescopes, with effective areas peaking at vastly different energies, are sensitive to different evaporation epochs; the terminal explosive, runaway phase is most constrained by the highest-energy, largest-effective-area observatories, such as the future CTA observatory; ongoing evaporation can be effectively searched for with GRB telescopes, including the Fermi Large Area Telescope.

While no evidence of exploding PBHs has emerged so far, perhaps the most promising path ahead is to use the Interplanetary Gamma-Ray Burst Timing Network, as exploding PBHs leaving a visible imprint are slated to be much closer than, approximately, 1 pc, as opposed to cosmologically-distant GRBs.

\vspace{1cm}
{\footnotesize {\bf Acknowledgments.}
I am deeply grateful to Benjamin Lehmann, who has driven most of the the work discussed here. It is because of students like Ben that I love my job. I am also very grateful to the reviewer, Kaloian D. Lozanov, for providing numerous suggestions and corrections upon reviewing an earlier version of this paper. This material is based upon work supported in part by  the U.S. Department of Energy, grant number de-sc0010107}.

\bibliographystyle{unsrt}
\bibliography{authorsample.bib}

\begin{thebibliography}{10}

\bibitem{Baker:2022rkn}
Michael~J. Baker and Andrea Thamm.
\newblock {Black hole evaporation beyond the Standard Model of particle
  physics}.
\newblock {\em JHEP}, 01:063, 2023.

\bibitem{Carr:2020gox}
Bernard Carr, Kazunori Kohri, Yuuiti Sendouda, and Jun'ichi Yokoyama.
\newblock {Constraints on primordial black holes}.
\newblock {\em Rept. Prog. Phys.}, 84(11):116902, 2021.

\bibitem{Lehmann_2019}
Benjamin~V. Lehmann, Christian Johnson, Stefano Profumo, and Thomas
  Schwemberger.
\newblock {Direct detection of primordial black hole relics as dark matter}.
\newblock {\em JCAP}, 10:046, 2019.

\bibitem{MacGibbon:1987my}
Jane~H. MacGibbon.
\newblock {Can Planck-mass relics of evaporating black holes close the
  universe?}
\newblock {\em Nature}, 329:308--309, 1987.

\bibitem{DeWitt:1975ys}
Bryce~S. DeWitt.
\newblock {Quantum Field Theory in Curved Space-Time}.
\newblock {\em Phys. Rept.}, 19:295--357, 1975.

\bibitem{Hawking:1976ja}
S.~W. Hawking.
\newblock {Zeta Function Regularization of Path Integrals in Curved
  Space-Time}.
\newblock {\em Commun. Math. Phys.}, 55:133, 1977.

\bibitem{Diamond:2021scl}
Melissa~D. Diamond and David~E. Kaplan.
\newblock {Constraints on relic magnetic black holes}.
\newblock {\em JHEP}, 03:157, 2022.

\bibitem{Page:1976ki}
Don~N. Page.
\newblock {Particle Emission Rates from a Black Hole. 2. Massless Particles
  from a Rotating Hole}.
\newblock {\em Phys. Rev. D}, 14:3260--3273, 1976.

\bibitem{Fujita:2014hha}
Tomohiro Fujita, Masahiro Kawasaki, Keisuke Harigaya, and Ryo Matsuda.
\newblock {Baryon asymmetry, dark matter, and density perturbation from
  primordial black holes}.
\newblock {\em Phys. Rev. D}, 89(10):103501, 2014.

\bibitem{Morrison:2018xla}
Logan Morrison, Stefano Profumo, and Yan Yu.
\newblock {Melanopogenesis: Dark Matter of (almost) any Mass and Baryonic
  Matter from the Evaporation of Primordial Black Holes weighing a Ton (or
  less)}.
\newblock {\em JCAP}, 05:005, 2019.

\bibitem{Viel:2005qj}
Matteo Viel, Julien Lesgourgues, Martin~G. Haehnelt, Sabino Matarrese, and
  Antonio Riotto.
\newblock {Constraining warm dark matter candidates including sterile neutrinos
  and light gravitinos with WMAP and the Lyman-alpha forest}.
\newblock {\em Phys. Rev. D}, 71:063534, 2005.

\bibitem{Lehmann:2021ijf}
Benjamin~V. Lehmann and Stefano Profumo.
\newblock {Black hole remnants are not too fast to be dark matter}.
\newblock {\em Phys. Dark Univ.}, 39:101145, 2023.

\bibitem{Page:1977um}
Don~N. Page.
\newblock {Particle Emission Rates from a Black Hole. 3. Charged Leptons from a
  Nonrotating Hole}.
\newblock {\em Phys. Rev. D}, 16:2402--2411, 1977.

\bibitem{1964ApJS....9..185B}
William~J. {Boardman}.
\newblock {The Radiative Recombination Coefficients of the Hydrogen Atom.}
\newblock {\em \apjs}, 9:185, August 1964.

\bibitem{Tretyak:2009sr}
V.~I. Tretyak.
\newblock {Semi-empirical calculation of quenching factors for ions in
  scintillators}.
\newblock {\em Astropart. Phys.}, 33:40--53, 2010.

\bibitem{Mu:2013pja}
Wei Mu and Xiangdong Ji.
\newblock {Ionization Yield from Nuclear Recoils in Liquid-Xenon Dark Matter
  Detection}.
\newblock {\em Astropart. Phys.}, 62:108--114, 2015.

\bibitem{Lazanu:2020qod}
Ionel Lazanu, Sorina Lazanu, and Mihaela P\^arvu.
\newblock {About detecting very low mass black holes in LAr detectors}.
\newblock {\em JCAP}, 10:046, 2020.

\bibitem{PICO:2016kso}
C.~Amole et~al.
\newblock {Improved dark matter search results from PICO-2L Run 2}.
\newblock {\em Phys. Rev. D}, 93(6):061101, 2016.

\bibitem{HiRes:2004hvo}
R.~U. Abbasi et~al.
\newblock {A Study of the composition of ultrahigh energy cosmic rays using the
  High Resolution Fly's Eye}.
\newblock {\em Astrophys. J.}, 622:910--926, 2005.

\bibitem{Aprile:2018dbl}
E.~Aprile et~al.
\newblock {Dark Matter Search Results from a One Ton-Year Exposure of XENON1T}.
\newblock {\em Phys. Rev. Lett.}, 121(11):111302, 2018.

\bibitem{Ghosh_1990}
D.~Ghosh and S.~Chatterjea.
\newblock {Supermassive magnetic monopoles flux from the oldest mica samples}.
\newblock {\em EPL}, 12:25--28, 1990.

\bibitem{Barrau:2019cuo}
Aur\'elien Barrau, Killian Martineau, Flora Moulin, and Jean-Fr\'ed\'eric
  Ngono.
\newblock {Dark matter as Planck relics without too exotic hypotheses}.
\newblock {\em Phys. Rev. D}, 100(12):123505, 2019.

\bibitem{Pilaftsis:2009pk}
Apostolos Pilaftsis.
\newblock {The Little Review on Leptogenesis}.
\newblock {\em J. Phys. Conf. Ser.}, 171:012017, 2009.

\bibitem{Smyth:2021lkn}
Nolan Smyth, Lillian Santos-Olmsted, and Stefano Profumo.
\newblock {Gravitational baryogenesis and dark matter from light black holes}.
\newblock {\em JCAP}, 03(03):013, 2022.

\bibitem{Dolgov:2011cq}
Alexander~D. Dolgov and Damian Ejlli.
\newblock {Relic gravitational waves from light primordial black holes}.
\newblock {\em Phys. Rev. D}, 84:024028, 2011.

\bibitem{Ireland:2023avg}
Aurora Ireland, Stefano Profumo, and Jordan Scharnhorst.
\newblock {Primordial gravitational waves from black hole evaporation in
  standard and nonstandard cosmologies}.
\newblock {\em Phys. Rev. D}, 107(10):104021, 2023.

\bibitem{Ireland:2023zrd}
Aurora Ireland, Stefano Profumo, and Jordan Scharnhorst.
\newblock {Gravitational Waves from Primordial Black Hole Evaporation with
  Large Extra Dimensions}.
\newblock 12 2023.

\bibitem{Boluna:2023jlo}
Xavier Boluna, Stefano Profumo, Juliette Bl\'e, and Dana Hennings.
\newblock {Searching for Exploding black holes}.
\newblock {\em JCAP}, 04:024, 2024.

\bibitem{Fermi-LAT:2017opo}
M.~Ackermann et~al.
\newblock {The Fermi Galactic Center GeV Excess and Implications for Dark
  Matter}.
\newblock {\em Astrophys. J.}, 840(1):43, 2017.

\bibitem{Carlson:2014ssa}
Eric Carlson, Adam Coogan, Tim Linden, Stefano Profumo, Alejandro Ibarra, and
  Sebastian Wild.
\newblock {Antihelium from Dark Matter}.
\newblock {\em Phys. Rev. D}, 89(7):076005, 2014.

\bibitem{Coogan:2017pwt}
Adam Coogan and Stefano Profumo.
\newblock {Origin of the tentative AMS antihelium events}.
\newblock {\em Phys. Rev. D}, 96(8):083020, 2017.

\bibitem{Cui:2016ppb}
Ming-Yang Cui, Qiang Yuan, Yue-Lin~Sming Tsai, and Yi-Zhong Fan.
\newblock {Possible dark matter annihilation signal in the AMS-02 antiproton
  data}.
\newblock {\em Phys. Rev. Lett.}, 118(19):191101, 2017.

\bibitem{Hawking:1974rv}
S.~W. Hawking.
\newblock {Black hole explosions}.
\newblock {\em Nature}, 248:30--31, 1974.

\bibitem{vonKienlin:2020xvz}
A.~von Kienlin et~al.
\newblock {The Fourth Fermi-GBM Gamma-Ray Burst Catalog: A Decade of Data}.
\newblock {\em Astrophys. J.}, 893:46, 2020.

\bibitem{Fermi-LAT:2021ykq}
L.~Baldini et~al.
\newblock {Catalog of Long-term Transient Sources in the First 10 yr of
  Fermi-LAT Data}.
\newblock {\em Astrophys. J. Supp.}, 256(1):13, 2021.

\bibitem{CTAConsortium:2013ofs}
B.~S. Acharya et~al.
\newblock {Introducing the CTA concept}.
\newblock {\em Astropart. Phys.}, 43:3--18, 2013.

\bibitem{Ukwatta:2015mfb}
T.~N. Ukwatta et~al.
\newblock {Investigation of Primordial Black Hole Bursts using Interplanetary
  Network Gamma-ray Bursts}.
\newblock {\em Astrophys. J.}, 826(1):98, 2016.

\bibitem{HAWC:2011gts}
A.~U. Abeysekara et~al.
\newblock {On the sensitivity of the HAWC observatory to gamma-ray bursts}.
\newblock {\em Astropart. Phys.}, 35:641--650, 2012.

\bibitem{Hamaide:2023ayu}
Louis Hamaide, Lucien Heurtier, Shi-Qian Hu, and Andrew Cheek.
\newblock {Primordial Black Holes Are True Vacuum Nurseries}.
\newblock 11 2023.

\bibitem{Hidalgo:2011fj}
J.~C. Hidalgo, L.~Arturo Urena-Lopez, and Andrew~R. Liddle.
\newblock {Unification models with reheating via Primordial Black Holes}.
\newblock {\em Phys. Rev. D}, 85:044055, 2012.

\end{thebibliography}

\end{document}